\documentclass[aps,pra,twocolumn,english,showpacs]{revtex4-1}
\usepackage{graphicx}
\usepackage{amssymb}
\usepackage{amsmath, amstext, amssymb, amsfonts, amsxtra, mathtools, stmaryrd}
\usepackage{braket}
\usepackage{textcomp}
\usepackage{relsize}
\usepackage[usenames,dvipsnames]{color}
\usepackage{units}
\usepackage[normalem]{ulem}
\newcommand{\aver}[1]{\langle #1 \rangle}

\begin{document}
\title{Cavity-induced generation of non-trivial topological states in a two-dimensional Fermi gas} 

\author{ Ameneh Sheikhan$^1$, Ferdinand Brennecke$^2$, and Corinna Kollath$^1$}

\affiliation{$^1$HISKP, University of Bonn, Nussallee 14-16, 53115 Bonn, Germany\\
$^2$ Physikalisches Institut, University of Bonn, Wegelerstr.~8, 53115 Bonn, Germany}
\begin{abstract}
We propose how topologically non-trivial states can dynamically organize in a fermionic quantum gas which is confined to a two-dimensional optical lattice potential and coupled to the field of an optical cavity. The spontaneously emerging cavity field induces together with coherent pump laser fields a dynamical gauge field for the atoms. Upon adiabatic elimination of the cavity degree of freedom, the system is described by an effective Hofstadter model with a self-consistency condition which determines the tunneling amplitude along the cavity direction. The fermions are found to self-organize into topologically non-trivial states which carry an extended edge state for a finite system size. Due to the dissipative nature of the cavity field, the topological steady states are protected from external perturbations.
\end{abstract}
\maketitle
The interest in topologically non-trivial quantum phases has revived recently with the discovery of topological insulators \cite{XiaoNiu2010}. In such materials extended edge states exist which are linked to the topological characteristics of the bulk  by the bulk-edge correspondence \cite{Hatsugai1993,WenHatsugai1994}. Since these edge states are non-local they are well protected against many environmental perturbations and therefore might be utilized for quantum computation \cite{NayakSarma2008}.

The preparation and observation of topologically generated edge states by static gauge fields has been achieved by now in several different systems. These reach from quantum Hall systems \cite{AndoUemura1975,Thouless1983,WakabayashiKawaji1978} over the more recent experiments in hybrid superconductor-semiconductor nanowires \cite{MourikKouwenhoven2012} or ferromagnetic atomic chains on a superconductor \cite{NadjPergeYazdani2014} and photonic materials \cite{HafeziTaylor2013} to ultracold neutral quantum gases \cite{LederWeitz2016}.

A novel approach for realizing edge states in dissipative systems has been proposed in Refs.~\cite{DiehlZoller2011,BardynImamoglu2012,BardynImamoglu2012b,IeminiMazza2016}
by means of engineering a tailored coupling to an environment. The dissipation results in an attractor dynamics \cite{MuellerZoller2012} which drives the system towards a steady state with the desired properties such as the existence of extended edge states. By the dissipative dynamics the state can be stabilized from destructive influences of its environment.

A natural dissipation channel is realized in ultracold quantum gases by coupling the atomic motion to the field of an optical cavity \cite{RitschEsslinger2013}. In the past, various self-organization phenomena in such systems have been investigated both theoretically and experimentally such as the Dicke phase transition \cite{DomokosRitsch2002,NagyDomokos2010,BaumannEsslinger2010,PiazzaZwerger2013} or more complex quantum phases reaching from phases in extended Bose-Hubbard models \cite{LarsonLewenstein2008,MaschlerRitsch2008,LandigEsslinger2016,KlinderHemmerich2015b,ChenZhai2016,DograDonner2016,Caballero-Benitez2016} over fermionic phases \cite{LarsonLewenstein2008b,MuellerSachdev2012,PiazzaStrack2014,KeelingSimons2014,ChenZhai2014}, magnetic phases \cite{GelhausenStrack2016}, phases in multimode cavities\cite{GopalakrishnanGoldbart2009,KollarLev2016} and disordered structures \cite{StrackSachdev2011,GopalakrishnanGoldbart2011,HabibianMorigi2013} to phases with spin-orbit coupling \cite{DengYi2014,DongPu2014,PanGuo2015,PadhiGosh2014}. More recently, the generation of a chiral current in ladder structures \cite{KollathBrennecke2016,SheikhanKollath2016,WolffKollath2016} and a transient current in a one-dimensional chain \cite{ZhengCooper2016} via the coupling of the tunneling dynamics to a cavity have been studied. In this paper we propose a scheme for the self-organization of an edge state in a two-dimensional ultracold Fermi gas loaded into a tilted optical lattice potential and coupled to the field of an optical cavity. The atoms are driven by coherent pump laser fields and scatter photons into the cavity field conditioned on a tunneling event between neighboring lattice sites. Above a critical pump strength a coherent cavity field emerges and the fermionic atoms experience a dynamical artificial gauge field in which they aquire topologically non-trivial properties. The {\it dynamical} gauge field realized in this work by the coupling to the cavity field has to be contrasted with the {\it static} artificial gauge fields, which have been induced recently in optical lattices thereby realizing the Hofstadter model \cite{JakschZoller2003, AidelsburgerBloch2013,MiyakeKetterle2013,AidelsburgerGoldman2015,AtalaBloch2014} and the Haldane model \cite{JotsuEsslinger2014}.
\begin{figure}
  \includegraphics[width=1\linewidth]{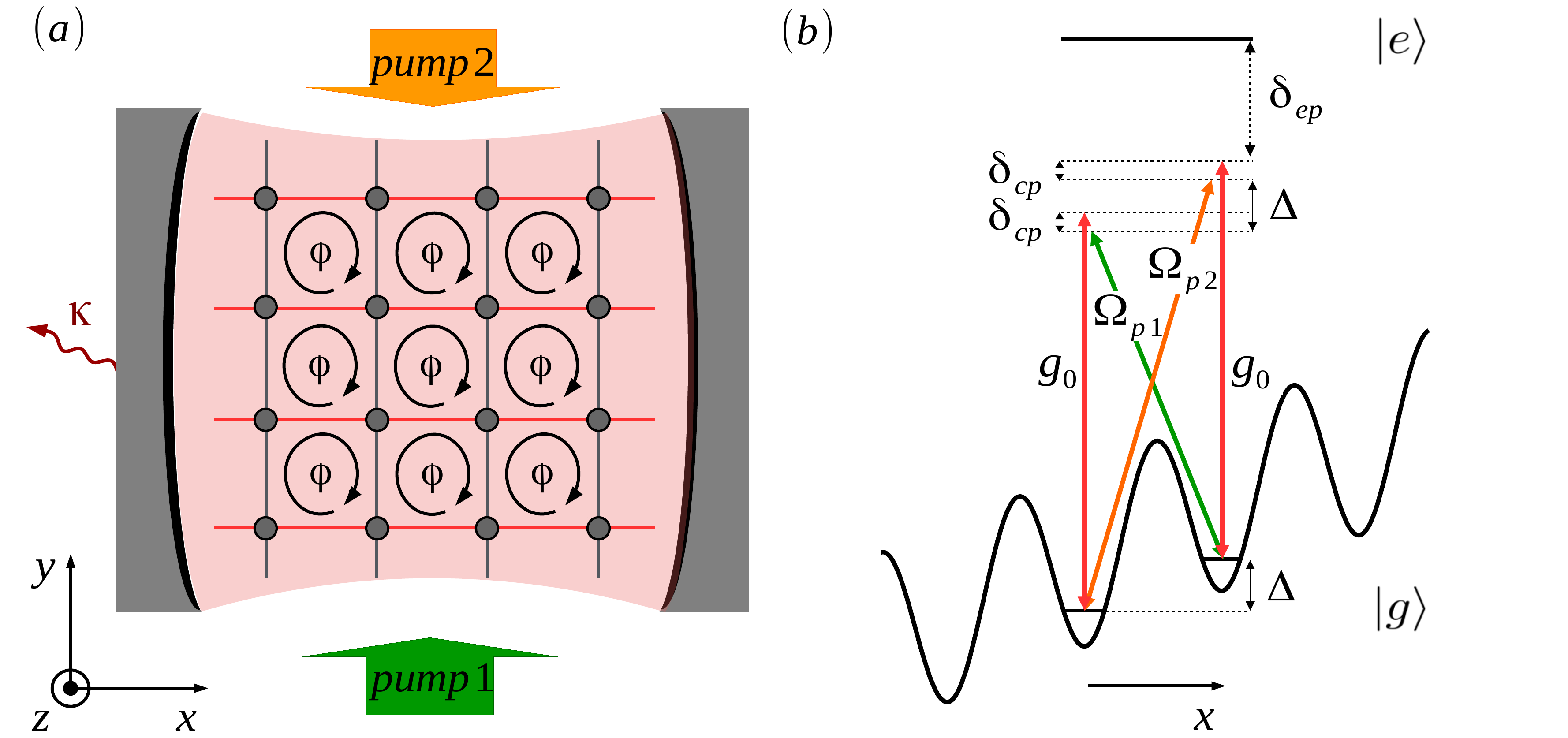}\\
  \includegraphics[width=1\linewidth]{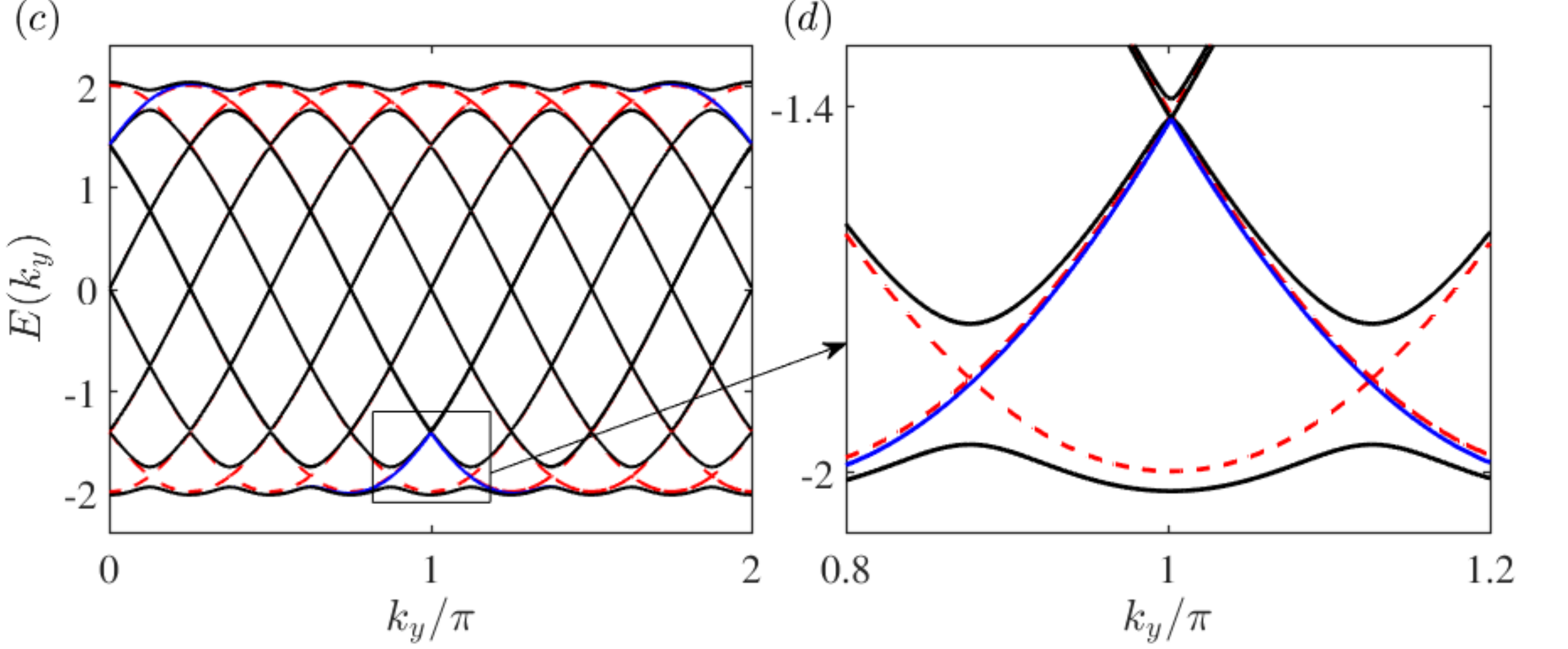}
\caption{\label{fig:setup}(color online) Scheme for generating a dynamical version of the Hofstadter model in a 2D optical lattice. 
(a) Fermionic atoms are loaded into a square optical lattice potential and coupled to the dynamical field of an optical cavity. Tunneling along the $x$-direction is strongly suppressed by a potential offset $\Delta$ between neighboring lattice sites. (b) Cavity-assisted Raman processes induced by two running-wave pump beams restore tunneling along the cavity direction and imprint a phase $\varphi$ onto the atomic wavefunction when tunneling around a plaquette. (c) The single-particle energy bands for the Hofstadter model of size $L_x=159, L_y=1000$ with periodic boundary condition in $y$-direction and open boundary condition in $x$-direction. The flux is chosen as $\varphi=2\pi \frac{1}{8}$. The solid (black) line shows the energy bands for $J_x/J_y=0.1$ and the (red) dashed lines correspond to decoupled chains with $J_x=0$. (d) Zoom into the first energy gap where the edge state (blue line) between two bulk bands for $J_x/J_y=0.1$ becomes visible.} 
\end{figure}

In our proposal we consider ultracold spin-polarized fermionic atoms confined to two spatial dimensions by a strong optical lattice potential along the $z$-direction (see Fig.~1a). Within the $x$-$y$-plane an additional optical square lattice potential is created  using standing-wave laser beams with wavelength $\lambda_x$ and $\lambda_y$ along the $x$-and $y$-direction, respectively. Along the $x$-direction a magnetic field gradient introduces a potential offset $\Delta$ between neighboring lattice sites thereby suppressing tunneling along this direction \footnote{An alternative route is to use a superlattice structure as realized in Ref.~\cite{AidelsburgerGoldman2015}.}. The fermions are placed into a high-finesse optical standing-wave cavity oriented along the $x$-direction and driven by two far-detuned, running-wave pump laser fields counter-propagating transversally to the cavity direction. Cavity-assisted tunneling processes along the $x$-direction are induced by photon exchange between either of the pump fields with frequencies $\omega_{p1,2}$ and a near-resonant cavity mode with resonance frequency $\tilde{\omega}_c \approx \omega_{p1} + \Delta/\hbar \approx \omega_{p2} - \Delta/\hbar$, see Fig.~1b. In order to selectively drive the indicated Raman transitions, the linewidth of the cavity is assumed to be on the order of the potential offset $\Delta/\hbar$.

We assume the frequencies of the cavity and pump fields to be far detuned from the atomic transition frequency $\omega_e$, such that the electronically excited state $|e \rangle$ of the atoms is only weakly occupied and can be adiabatically eliminated. This allows an effective description of the system dynamics in terms of the dynamical cavity field and the atomic ground state $|g\rangle$, expanded in the Wannier basis of the optical lattice potential using the tight-binding approximation \cite{MaschlerRitsch2008, RitschEsslinger2013}. Neglecting the AC-Stark shift induced by the intra-cavity photons and off-resonant two-photon transitions, the Hamiltonian becomes
\begin{eqnarray}
H_y&&=-J_y \sum_{j,m} \left( c^\dagger_{m,j} c_{m,j+1}+\mathrm{H.c.}\right),\, H_c=\hbar \delta_{cp} a^\dagger a,\nonumber\\
\textrm{and} \,\,&& H_{ac} = - \hbar\tilde{\Omega} (a^\dagger+a) \sum_{m,j} \left (e^{i \varphi j} c^\dagger_{m,j} c_{m+1,j}+\mathrm{H.c.} \right ).\nonumber
\end{eqnarray}
Here, $c_{m,j}$ $(c_{m,j}^{\dag})$ are the annihilation (creation) operators of an atom on lattice site $(m,j)$, where $m$ ($j$) enumerates the lattice sites along the $x$ ($y$)-direction, respectively. The term $H_y$ describes tunneling of the atoms along the $y$-direction with amplitude $J_y$. The term $H_c$ captures the bare dynamics of the dispersively shifted cavity field in a frame rotating at frequency $\omega_p$ where we denoted the average pump-cavity detuning by $\delta_{cp}=\left(\tilde{\omega}_c-\omega_p \right)$ with $\omega_p=\frac{1}{2}(\omega_{p1}+\omega_{p2})$. The operator $a$ ($a^\dag$) denotes the annihilation (creation) operator of a cavity photon.
The cavity-assisted tunneling processes of atoms along the $x$-direction are described by the term $H_{ac}$. The coupling strength between the atomic tunneling operator and the cavity field is given by $\hbar\tilde{\Omega}=\frac{\hbar\Omega_{p1} g_0}{\omega_e-\omega_{p1}}\phi_{_\parallel}\phi_{\perp}$, where $\Omega_{p1}$ denotes the Rabi frequency of the first pump beam and $g_0$ the vacuum-Rabi frequency of the cavity. The Rabi frequency of the second pump beam is chosen as $ \Omega_{p2}=\frac{\Omega_{p1} (\omega_e-\omega_{p2})}{(\omega_e-\omega_{p1})} $ in order to balance the strength of the two Raman channels. The wave length of the cavity is chosen $\lambda_c=\frac{\lambda_x}{2}$. The overlap integrals $\phi_\parallel$ and $\phi_{\perp}$ are effective parameters depending on the Wannier states and can be tuned via the geometry of the optical lattice and the cavity mode \cite{KollathBrennecke2016}.

The running-wave character of the transverse pump fields is utilized to imprint during a cavity-assisted tunneling event the phase factor $\exp(i k_p y/a_y)$ onto the atomic wavefunction. Here, $k_p = 2\pi a_y/\lambda_p$ denotes the unit-less pump wave vector with $\lambda_p\approx \lambda_{p1}\approx \lambda_{p2}$. $a_y$ is the lattice constant in $y$-direction. Correspondingly, if atoms tunnel around a plaquette of the square lattice, they collect a total phase of $\varphi = \pi \lambda_y/\lambda_p$\footnote {Other contributions to the phase factor exist. However, since these do not contribute to the phase collected while tunneling around a plaquette, they are not detailed here.}. This corresponds for the atoms to the presence of a (dynamical) artificial magnetic field oriented along the $z$-direction. Its magnitude can be tuned by changing  the $y$-component of the pump wavevector.

The loss of intracavity photons through the imperfect cavity mirrors can be theoretically described in terms of a Lindblad equation
$\frac{\partial}{\partial t} \rho= -i/\hbar [H,\rho]+\mathcal{D}(\rho)$ where $\rho$ denotes the density matrix of the combined atoms-cavity system. The dissipation of cavity photons is captured by the term $\mathcal{D}(\rho)=\kappa\left(2 a \rho  a^\dagger -(a^\dagger a \rho+\rho a^\dagger a)\right)$ with cavity decay rate $\kappa$. Solving the above model exactly is very complicated. Therefore we concentrate on finding the steady states of the system  using adiabatic elimination of the cavity field. Based on our earlier work \cite{KollathBrennecke2016, WolffKollath2016} we expect this method to cover most of the important physical properties of the steady states.

After solving the equations of motion for the cavity field expectation value $\alpha=\aver{a}= \frac{\tilde{\Omega}}{(\delta_{cp}-i \kappa)}\aver{  K_x+K_x^\dagger}$ and substituting it on a mean-field level into the equations of motion of the atomic operators one arrives at the following effective Hamiltonian 
\begin{eqnarray}
H_{F}=H_y - (J_x K_x+\mathrm{H.c.})
\label{eq:Heff}
\end{eqnarray}
together with the self-consistency condition $J_x= A \aver{K_x+K_x^\dagger}/2$ where $A= \frac{4 \hbar \tilde{\Omega}^2 \delta_{cp}}{\delta_{cp}^2+ \kappa^2}$. Here, $K_x =\sum_{m,j}  e^{i \varphi j}  c^\dagger_{m,j} c_{m+1,j}$ denotes the directed atomic tunneling operator along the $x$-direction.  The $Z_2$-symmetry, corresponding to the system's invariance under the transformation $a \rightarrow -a$ and $c_{m,j}\rightarrow (-1)^{m} c_{m,j}$, is reflected in the freedom of the sign of the cavity field expectation value $\alpha$ and of the expectation value of the tunneling along the $x$-direction $\aver{K_x+K_x^\dagger}$. In an experiment this symmetry will be spontaneously broken by the cavity dissipation and in what follows we choose $\aver{K_x+K_x^\dagger}>0$. Since for the ground state of $H_{F}$ the expectation value of the directed tunneling $\aver{K_x+K_x^\dagger}$ has the same sign as $J_x$, a non-trivial solution of the self-consistency condition only exists for a positive pump-cavity detuning $\delta_{cp}>0$. A solution of the effective model with $J_x>0$ corresponds to a finite expectation value of the cavity field operator given by $\alpha= J_x(\delta_{cp}+i \kappa)/(2\hbar \tilde{\Omega} \delta_{cp})$.

We numerically solve the self-consistent problem of the effective Hamiltonian $H_F$ taking into account periodic boundary conditions in the $y$-direction with number of sites $L_y$ and open boundary conditions along the $x$-direction with number of sites $L_x$. We choose the flux to be $\varphi=2\pi\frac{p}{q}$ where $p$ and $q$ are mutually prime. The number of sites along the $x$-direction is taken to be commensurate with the flux, i.e.~as $L_x=lq-1$ with $l$ being an integer. In this geometry the total number of plaquettes is $N_p=L_y(L_x-1)$ and two edges arise at $m=1$ and $m=L_x$. In order to diagonalize the Hamiltonian $H_F$ we use the gauge transformation $c_{m,j}=\tilde{c}_{m,j}e^{-i\varphi m j}$ and  decouple different momenta along the $y$-direction using the momentum representation $\tilde{c}_{m,k_y}=\frac{1}{L_y}\sum_j e^{-i k_y j} \tilde{c}_{m,j}$ to determine the single-particle eigenenergies of the Hamiltonian for each momentum $k_y$ independently.

\begin{figure}
\includegraphics[width=0.99\linewidth]{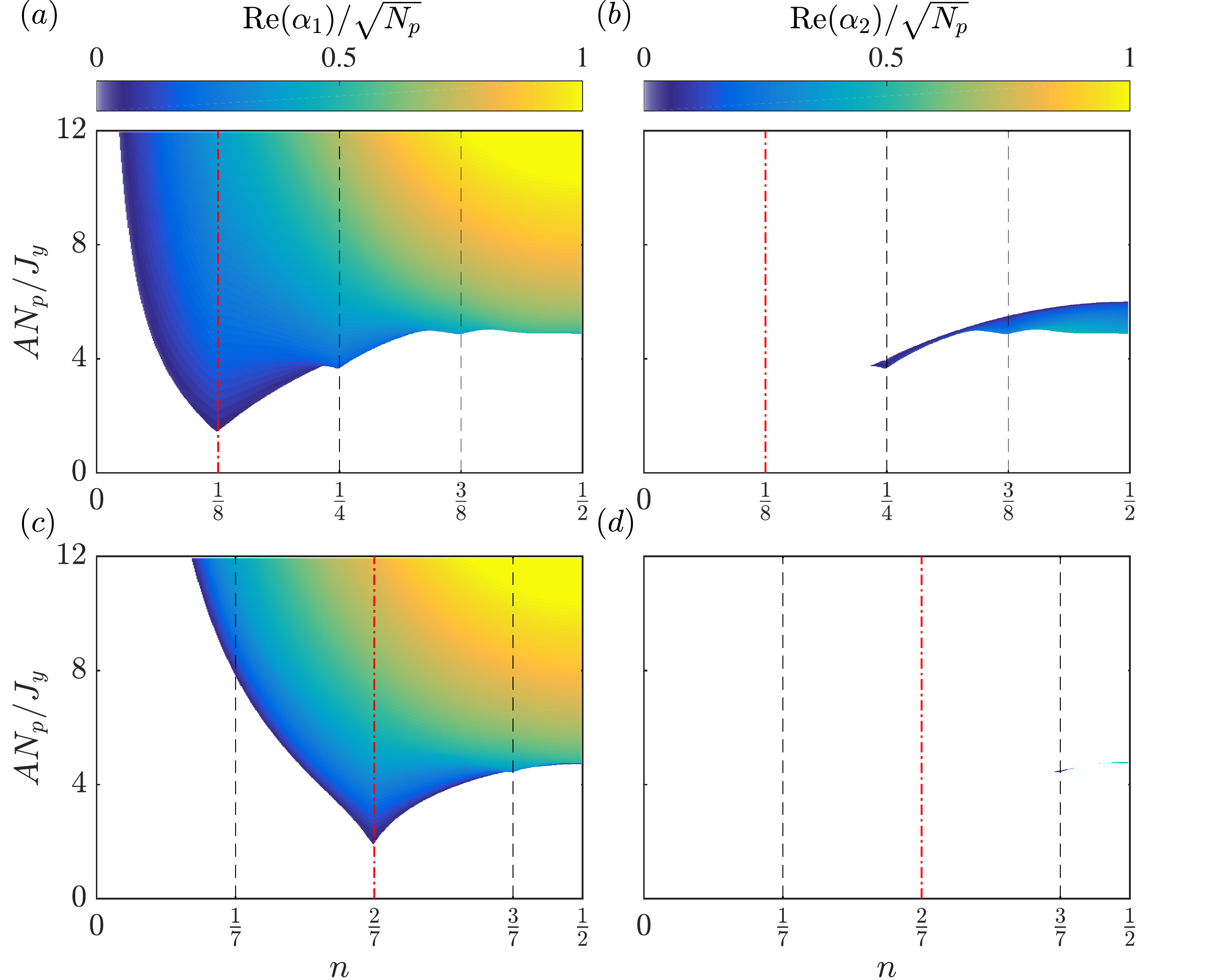}
\caption{\label{fig:field}(color online) Steady-state amplitude of the cavity field ${\text {Re}}(\alpha)/\sqrt{N_p}$ (color code) for flux $\varphi=2\pi\frac{1}{8}$ (a,b) and $\varphi=2\pi\frac{2}{7}$ (c,d) as a function of filling $n$ and  pump strength $A N_p/J_y$. Shown are the first (left panels) and the second (right panels) solution of the effective model Eq.1 using adiabatic elimination of the cavity field. White regions indicate the trivial empty cavity solution. The dashed-dotted, red lines correspond to the filling where the Fermi energy lies inside the gap with $2\pi n=\varphi$. The dashed, black lines indicate the fillings where the Fermi energy lies inside higher-order gaps. Here, $L_x=159$, $L_y=1000$, $\hbar\delta_{cp}=J_y$ and $\hbar\kappa = 0.05 J_y$.}
\end{figure}

In Figs.~\ref{fig:field}, we show the steady-state cavity field amplitude $Re(\alpha)/\sqrt{N_p}$ versus the filling $n = \frac{1}{L_x L_y}\sum_{m,j}\aver{c^\dagger_{m,j} c_{m,j}}$ and pump strength $A$ for two different values of the flux. Since the behavior of the system is symmetric around half-filling, we only show fillings below $n\le 1/2$.
 For both fluxes we find over a large range of fillings a critical value of the pump strength $A_c$ above which a finite cavity field amplitude builds up. 
In a certain parameter regime we find the coexistence of two steady-state solutions as shown in  Fig.~\ref{fig:field} (a) and (b) for $\varphi=2\pi\frac{1}{8}$. For flux $\varphi=2\pi\frac{2}{7}$ there is a very small region of parameters within that the second steady-state solution exists (Fig.~\ref{fig:field} (d)). To decide on the stability of the second solution, which we found in a finite system to be unstable, a stability analysis should be performed. In the following we concentrate on the first solutions shown in Fig.~\ref{fig:field} (a) and (c).

A special situation arises for the filling  $n=\frac{p}{q}$ where the Fermi energy lies at the crossing point of the energy bands of uncoupled ($J_x=0$) neighbouring chains (chains are taken along the $y$-direction). In this situation, the cavity-assisted tunneling process of an atom between neighboring chains is a resonant process. Due to this resonance condition it is favourable for the cavity field to become occupied even for an infinitesimally small intensity of the pump beams. Thus, in this case, the critical pump strength vanishes and a continuous rise of the field amplitude with increasing pump strength is observed. A logarithmic onset of the tunneling expectation value $\aver{K_x+K_x^\dagger}$ along the $x$-direction in the Hofstadter model $H_F$ as a function of the tunneling amplitude $J_x$ is found, i.e.~$\aver{K_x+K_x^\dagger}\propto - J_x \log (J_x)$ (see Appendix \ref{app:Kperp}). Also around other fillings (marked with dashed lines in Fig.~\ref{fig:field}) where the Fermi energy hits the crossing point between energy bands of more distant chains, a higher-order resonant tunneling process couples these chains. Correspondingly, characteristic dips of the critical pump strength occur at these fillings, however, compared to the dip at filling $n=\frac{p}{q}$, they are typically much less pronounced and in some cases even invisible in the Fig.~\ref{fig:field}.

\begin{figure}
  \includegraphics[width=0.99\linewidth]{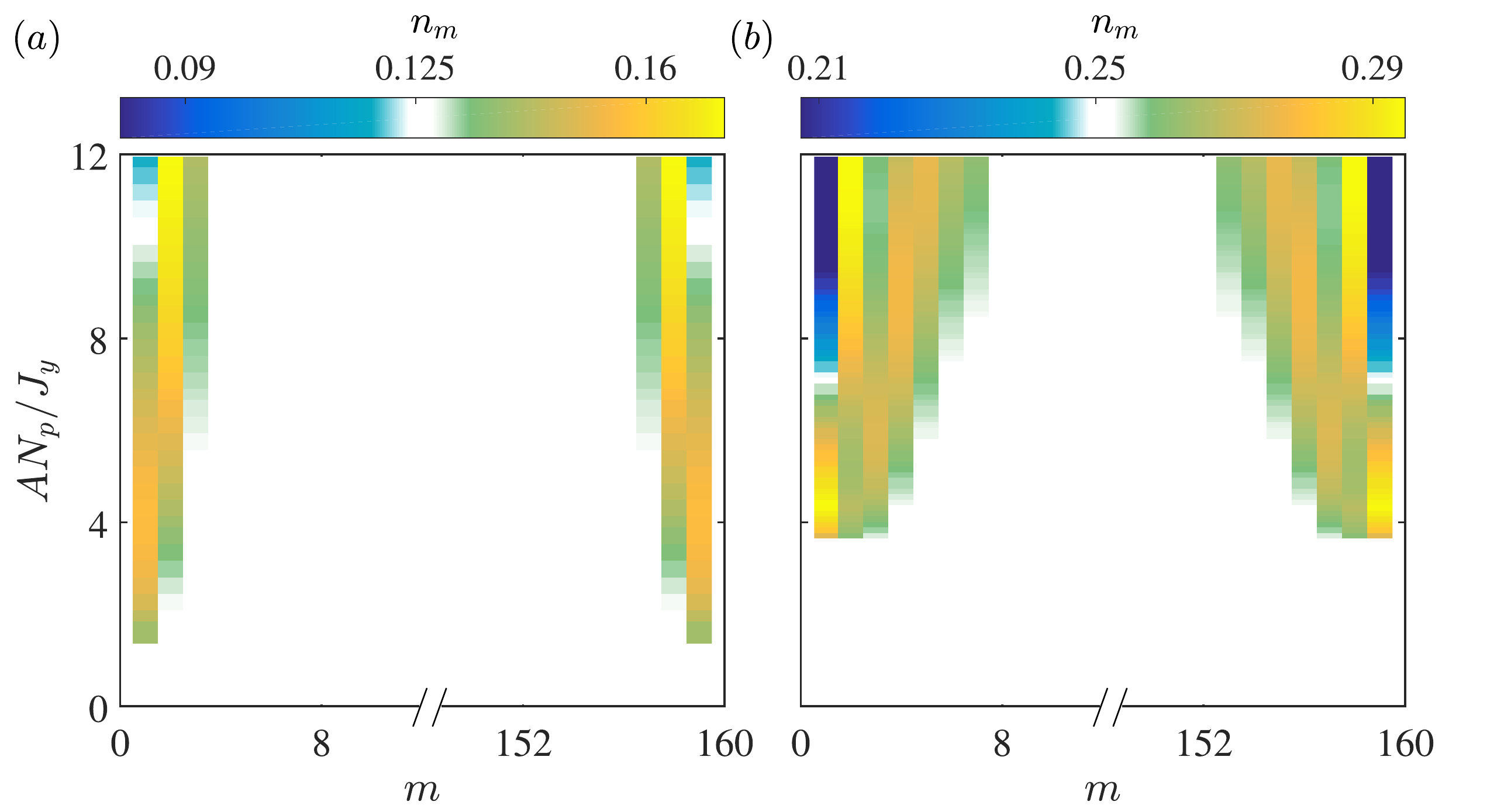}\\
 \caption{\label{fig:edge}(color online) Atomic density profile along the cavity direction in the steady state plotted versus pump strength $A N_p/J_y$ for a flux value of $\varphi=2\pi\frac{1}{8}$. A clear formation of an edge state is visible. The filling is chosen to yield a Fermi energy (a) inside the first gap with $n=0.126$ or (b) inside the second gap with $n=0.251$.}
\end{figure}

At the boundaries of topologically non-trivial phases (as they occur e.g.~in the static Hofstadter model) and topologically trivial phases (as the vacuum) an edge state forms. The energy of the edge state lies in between the bulk energy bands (see Fig.~\ref{fig:setup} (c) and (d)). In the considered setup we find the dynamic organization of chiral edge states. They form for fillings where the Fermi energy lies inside the gap of the energy bands, either above a critical pump strength or at infinitesimally small pump strength for the filling $n=p/q$. This is exemplified in Fig.~\ref{fig:edge} where we show the dependence of the fermionic density profile $n_m = \frac{1}{L_y}\sum_{j}\aver{c^\dagger_{m,j} c_{m,j}}$ on the pump strength $A$ for two different fillings, one of them corresponding to the filling $n\approx p/q$ (Fig.~\ref{fig:edge} (a)). The formation of an edge mode which is localized to a few lattice sites is clearly visible in a sharp rise of the atomic density at the boundary of the system size. Thus, by the atom-cavity coupling an edge mode is dynamically stabilized by the feedback mechanism with the cavity field.

\begin{figure}
\includegraphics[width=.99\linewidth]{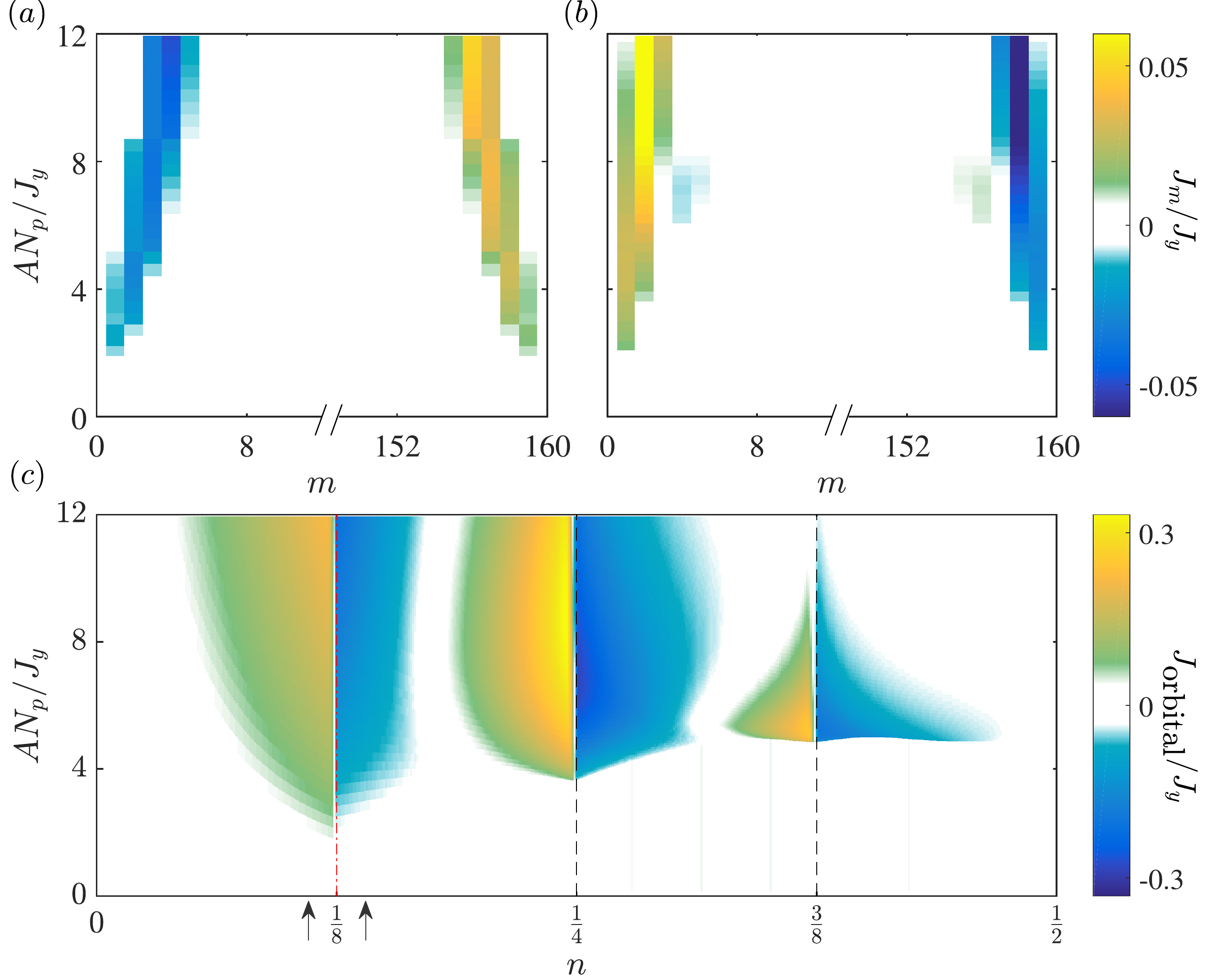}
\caption{\label{fig:Jcorbital}(color online)  The local current $J_m$ along the cavity direction in the steady state plotted versus pump strength $A N_p/J_y$ for a flux value of $\varphi=2\pi\frac{1}{8}$. The direction of the current is different for the fillings where the Fermi energy lies (a) below the first gap with $n=0.11$ and (b) above the first gap with $n=0.14$. (c) The orbital current in the steady state versus filling $n$ and pump strength $A N_p/J_y$ for the same flux. The orbital current jumps whenever the Fermi energy crosses a gap in the energy spectrum. The dashed-dotted, red line corresponds to the filling where the Fermi energy lies inside the gap with $2\pi n=\varphi$. The dashed (black) lines indicate the fillings where the Fermi energy lies inside higher-order gaps. The arrows indicate the fillings chosen in (a) and (b).}
\end{figure}

Topologically non-trivial phases are often connected to their chiral edge current.  The current along the legs is given by
\begin{eqnarray}
J_m=\frac{-i J_y}{L_y} \sum_{j} (c_{m,j}^\dagger c_{m,j+1}-c_{m,j+1}^\dagger c_{m,j}).\nonumber
\end{eqnarray}
In Fig.~\ref{fig:Jcorbital} we show the current $J_m$ versus the pump strength $A$ for fillings where the Fermi energy lies below (a) and above (b) the energy gap in order to make the contribution of the edge state visible. The current is localized at or close to the edges and can oscillate with the distance from the boundary. Its direction changes in between the two fillings displayed in Fig.~\ref{fig:Jcorbital}. We introduce  the orbital current \cite{CaioBhaseen2015} defined as 
\begin{eqnarray}
J_{\text{orbital}}=\frac{2}{L_x-1}\sum_{m=1}^{L_x}\left(\frac{L_x+1}{2}-m\right)J_m.\nonumber
\end{eqnarray}
This quantity measures the chiral current weighted by the distance from the center such that the current at the boundary is the most important one. The orbital current has been shown in Ref.~\cite{CaioBhaseen2015} to be a useful measure for the edge current. In Fig.~\ref{fig:Jcorbital} (c) the dependence of the orbital current on filling and pump strength is shown for a flux of $\varphi=2\pi \frac{1}{8}$. While the orbital current vanishes below the critical pump strength (cf.~Fig.~\ref{fig:field}), it typically takes a finite value above the critical pump strength.  Around the fillings where the Fermi energy lies within an energy gap, a jump signals the large contribution of the edge state which is another signature for the presence of the edge state.

An experimental detection of the edge modes could be indirectly performed by observing a non-zero cavity output field. Additionally, a more direct observation of the edge state via the fermionic density profile could be performed using high-resolution imaging techniques \cite{ParsonsGreiner2015,CheukZwierlein2015,HallerKuhr2015,OmranGross2015,BollGross2016,CocchiKoehl2016}.
\begin{acknowledgments}
We acknowledge fruitful discussions with M.~Fleischhauer, F.~Piazza, H.~Ritsch, and W.~Zwerger and support from the DFG through TR 185 and FOR 1807 and the ERC (Grant Number 648166, Phon(t)on). 
\end{acknowledgments}
\appendix
\section{Expectation value of the tunneling along $x$ direction}
\label{app:Kperp}
{In this section we calculate the expectation value $\aver{K_x+K_x^\dagger}$ of the tunneling along $x$-direction in the Hofstadter Hamiltonian $H_F$ (Eq.~\ref{eq:Heff}) for very small tunneling amplitudes $J_x\ll J_y$ at filling $n=\frac{p}{q}$. The knowledge about the behaviour of the tunneling expectation value  $\aver{K_x+K_x^\dagger}$ is required to solve the self-consistency condition and in particular in order to determine whether the continuous onset of the occupation of the cavity field is present at this filling. The filling  $n=\frac{p}{q}$ is particular, since at this filling for $J_x=0$ the Fermi energy lies at the crossing point of different energy bands and the splitting of these bands generate a logarithmic increase of the tunneling at small values of $J_x$.

In order to determine the main contribution of the bulk of the system to the tunneling expectation value $\aver{K_x+K_x^\dagger}$, we use the infinite Hofstadter model of Eq.~\ref{eq:Heff} in the momentum representation,
\begin{eqnarray}
H_F&=&\sum_{k_x,k_y^0}H_F(k_x,k_y^0)\nonumber\\
H_F(k_x,k_y^0)&=&-J_x\sum_{r=0}^{q-1} \left (e^{i k_x} c^\dagger_{k_x,k_y^0- r \varphi} c_{k_x,k_y^0- (r+1) \varphi}+\mathrm{H.c.} \right ) \nonumber\\
    &-&2J_y \sum_{r=0}^{q-1} \cos(k_y^0- r \varphi) c^\dagger_{k_x,k_y^0- r \varphi} c_{k_x,k_y^0- r \varphi}\nonumber
\label{eq:app_hamckxky}
\end{eqnarray}
where $c_{k_x,k_y}=\frac{1}{L_x L_y}\sum_{j,m} e^{-i (k_x m+k_y j)}c_{m,j}$ and $k_x\in [-\pi, \pi)$ and $k_y^0\in [-\frac{\pi}{q},\frac{\pi}{q})$.

From this form of the Hamiltonian (Eq.~\ref{eq:app_hamckxky}) one sees that always $q$ states are coupled. 
For $q=3$, for example, the Hamiltonian is given by
\begin{eqnarray}
\frac{H_F}{J_y}=
 \begin{pmatrix}
  -2 \cos (k_y^0+\frac{\varphi}{2}) &  -{\tilde J}& -\tilde J e^{-i q k_x}\\
  -{\tilde J}           &  -2 \cos (k_y^0 - \frac{\varphi}{2}) & -\tilde J\\
  -\tilde J e^{+i q k_x}  &  -\tilde J  & -2 \cos (k_y^0-\frac{3\varphi}{2}) 
 \end{pmatrix}\nonumber
\label{eq:app_hambulkq3}
\end{eqnarray}
where we shifted the momentum along $y$-direction by $k_y^0\rightarrow k_y^0+\frac{\varphi}{2}$ and $\tilde J\equiv\frac{J_x}{J_y}$. For ${\tilde J}= 0 $ , the energy of the first two eigenstates is degenerate at $k_y^0=0$ (for the considered shift) and the Fermi level at filling $n=\frac{1}{3}$ lies at this energy. 

We consider three cases, (i): $k_y^0\ll {\tilde J}\ll 1$, (ii): ${\tilde J}\ll k_y^0\ll 1$ and (iii): ${\tilde J}\ll 1 ~\&~ k_y^0\gg 1 $.
For case (i) the Hamiltonian is,

\begin{eqnarray}
\frac{H_F}{J_y}=
 \begin{pmatrix}
  \lambda                 &  -{\tilde J}  & -\tilde J e^{-i q k_x}\\
  -{\tilde J}             &  \lambda      & -\tilde J\\
  -\tilde J e^{+i q k_x}  &  -\tilde J    & \lambda_3
 \end{pmatrix}\nonumber
\end{eqnarray}
where $\lambda=-2\cos(\frac{\varphi}{2})$ and $\lambda_3=-2\cos(\frac{3\varphi}{2})$. 
 The eigenvalues are $\lambda'=\lambda\pm {\tilde J}$, $\lambda_3'=\lambda_3+\frac{2 \tilde J^2}{\lambda_3-\lambda}$ and the eigenvectors are $\frac{1}{\sqrt{2}}(1,1,0)$, $\frac{1}{\sqrt{2}}(1,-1,0)$ and $(0,0,1)$ up to the first order of magnitude in $\tilde J$. If only the lowest band is filled (which corresponds to $n=\frac{1}{3}$) the directed tunneling is $K_x(k_y^0\ll\tilde J)\approx\frac{1}{2}$.

For case (ii) the Hamiltonian is:
\begin{eqnarray}
\frac{H_F}{J_y}=
 \begin{pmatrix}
  \lambda-\eta k_y^0        &  -{\tilde J}           & -\tilde J e^{-i q k_x}\\
  -{\tilde J}             &  \lambda+\eta k_y^0      & -\tilde J\\
  -\tilde J e^{+i q k_x}  &  -\tilde J             & \lambda_3
 \end{pmatrix}\nonumber
\end{eqnarray}

where $\eta=2 \sin(\frac{\varphi}{2})$. The eigenvalues are $\lambda'=\lambda\pm(\eta k_y^0+\frac{\tilde J^2}{2\eta k_y^0})$ and $\lambda_3'=\lambda_3+\frac{2 \tilde J^2}{\lambda_3-\lambda}$. The eigenvectors are $(1,\frac{\tilde J}{2\eta k_y^0}, \frac{\tilde J e^{i q k_x}}{\lambda_3-\lambda})$, $(\frac{-\tilde J}{2\eta k_y^0},1,\frac{\tilde J}{\lambda_3-\lambda})$ and $(\frac{-\tilde J e^{-i q k_x}}{\lambda_3-\lambda},\frac{-\tilde J}{\lambda_3-\lambda},1)$. For this regime the directed tunneling is calculated as $K_x({\tilde J}\ll k_y^0\ll 1)\approx\frac{\tilde J}{2\eta k_y^0}+\frac{\tilde J}{\lambda_3-\lambda}$ which can be very large for very small momentum $k_y^0$. The directed tunneling in case (iii) is very small compared to the other cases. The same result is calculated for $\aver{K_x^\dagger}$.

Integrating the total directed tunneling over the Brillouin zone by dividing it into the three cases, results in a logarithmic behaviour of the tunneling along $x$-direction $\aver{K_x+K_x^\dagger}\propto -\tilde J \log (\tilde J)$ at small tunneling. We verified numerically that this behavior persists in the evaluation of integrals. Similar arguments can be repeated for different values of $q$, where subblocks of the matrices have a similar structure as the one discussed for $q=3$. Additionally, the behaviour of the edge states can be analyzed and leads to the same logarithmic dependence of the rung tunneling on the tunneling amplitudes. We tested numerically that up to large values of $q$, the described behaviour can be found both in the bulk and edge states. The consequence is that at filling $n=\frac{p}{q}$ an infinitesimally small pump strength is enough to stabilize the finite occupation of the cavity mode. 

}
%

\end{document}